\documentclass[prc,english,showpacs,showkeys,superscriptaddress,
nofootinbib,preprintnumbers,amsmath,amssymb,fleqn,byrevtex]{revtex4}
\usepackage{amsmath,amsfonts,amssymb,amscd,amsxtra,amsthm,amstext}
\usepackage[T1]{fontenc}
\usepackage[latin9]{inputenc}
\usepackage{bm}
\usepackage{graphicx}
\usepackage{esint}
\usepackage{dcolumn}
\usepackage{babel}
\usepackage{hyperref}


\begin{document}

\preprint{INHA-NTG-02/2011}

\title{Mass splittings of the baryon decuplet and antidecuplet with the
second-order flavor symmetry breakings within a chiral soliton model}

\author{Ghil-Seok Yang}
\email{ghsyang@gmail.com}
\affiliation{Center for High Energy Physics and Department of Physics,
  Kyungpook National University, Daegu 702-701, Republic of Korea} 

\author{Hyun-Chul Kim}
\email{hchkim@inha.ac.kr}
\affiliation{Department of Physics, Inha University, Incheon 402-751,
  Republic of Korea}
\affiliation{School of Physics, Korea Institute for Advanced Study,
  Seoul 130-722, Republic of Korea}

\date{June, 2012}
\begin{abstract}
We revisit the mass splittings of SU(3) baryons, taking into account
the second-order effects of isospin and SU(3) flavor symmetry
breakings within the framework of a chiral soliton model. The masses
of the baryon decuplet turn out to be improved, compared to those with
the first-order corrections. The mass of the $N^*$ as a member of the 
baryon antidecuplet is obtained as $M_{n^*}=1687$ MeV, which is
in agreement with the recent experimental data. The pion-nucleon sigma
term becomes $\Sigma_{\pi N}=(50.5\pm 5.4)$ MeV.
\end{abstract}

\pacs{12.39.Fe, 12.40.-y, 14.20.Dh}

\keywords{Mass splittings of SU(3) baryons, chiral soliton model,
  flavor symmetry breaking}

\maketitle

\section{Introduction}

The $\Theta^+$
baryon~\cite{Jezabek:1987ns,Diakonov:1997mm,Praszalowicz:2003ik}, 
which is the first excited exotic pentaquark state, has drawn much
attention, since the LEPS collaboration announced the evidence of its
existence ~\cite{Nakano:2003qx}. However, the null results of the CLAS
experiments about $\Theta^{+}$ 
~\cite{Battaglieri:2005er,McKinnon:2006zv,Niccolai:2006td,DeVita:2006ha},  
have cast doubt upon its existence. In the meanwhile, the DIANA
collaboration published the positive evidence of
$\Theta^{+}$~\cite{Barmin:2006we,Barmin:2009cz}. Very recently, 
it has announced the formation of a narrow $pK^{0}$ peak with mass of 
$1538\pm 2$ MeV/$c^{2}$ and width of $\Gamma=0.39\pm0.10$ MeV in the
$K^{+}n\rightarrow K^{0}p$ reaction with higher statistical
significance ($6\,\sigma-8\,\sigma$)~\cite{Barmin:2009cz}. In
addition, other new positive experiments for the $\Theta^{+}$ have been
reported~\cite{new_SVD,Hotta:2005rh,Miwa:2006if,SVD2:2008}.  The LEPS
collaboration reported again the evidence of the $\Theta^{+}$
existence~\cite{Nakano:2008ee} with the data
$M_{\Theta}=1.524\pm0.002\pm0.003\,\mathrm{GeV}/c^{2}$ given in the 
statistical significance $5.1\,\sigma$. 

In addition to $\Theta^+$, Kuznetsov et al.~\cite{Kuznetsov:2004gy,
  Kuznetsov:2006de, Kuznetsov:2006kt} discovered a new
nucleon-like resonance around $1.67$ GeV from $\eta$ 
photoproduction off the deutron in the neutron channel, based on the
GRAAL data. The decay width was measured to be around $40$
MeV. Excluding the effects of the Fermi-motion, Fix et
al.~\cite{Fix:2007st} argued that the width would further decrease.
A important point is that this resonant structure is only seen in the
neutron channel, which is the typical characteristics for the
photo-excitation of the non-strange antidecuplet
pentaquark~\cite{Polyakov:2003dx,Kim:2005gz}. Very recently, a new 
analysis of the free proton GRAAL data~\cite{Kuznetsov:2007dy,
  Kuznetsov:2008gm,Kuznetsov:2010as,Kuznetsov:2008hj,
  Kuznetsov:2008ii}  has revealed a resonance structure with a mass
around 1685 MeV and width $\Gamma\leq15$ MeV.
The CB-ELSA collaboration~\cite{CBELSA} has confirmed 
an evidence for this $N^{*}$ resonance. All these experimental facts
are consistent with the results for the transition magnetic moments in
the $\chi$QSM~\cite{Polyakov:2003dx,Kim:2005gz}  
and phenomenological analysis for the non-strange pentaquark
baryons~\cite{Azimov:2005jj}. 
Based on these results, theoretical calculations of the $\gamma
N\to\eta N$ reaction~\cite{Choi:2005ki,Choi:2007gy} were shown to
describe qualitatively well the GRAAL data. In
Refs.~\cite{Diakonov:2003jj,Arndt:2003ga,non_strange_partner2} 
the non-strange partners of the $\Theta^{+}$ were also studied.

In the present dubious situation related to the existence of
$\Theta^+$, one needs to carefully review the previous theoretical
analyses~\cite{Diakonov:1997mm,Ellis:2004uz} of the mass splittings of
SU(3) baryons. The original analysis~\cite{Diakonov:1997mm} was partially
based on specific model
calculations~\cite{Christov:1993ny,Blotz:1994wi}, while some dynamical 
parameters are fixed by some experimental masses of the baryon octet
and decuplet, and the empirical value of the $\pi N$ 
sigma term $\Sigma_{\pi N}$. Moreover, Diakonov et
al.~\cite{Diakonov:1997mm} assumed then $N^*(1710)$ to be a member of
the antidecuplet. Furthermore, the second moment of
inertia is an essential quantity in determining the shift of
the antidecuplet center from the octet center in the chiral limit but 
is known only in a wide range: $0.43\,\mathrm{fm}\,<\,
I_{2}\,<0.55\,\mathrm{fm}$, depending on specific models such as
either the Skyrme model~\cite{Walliser:1992am,Walliser:1992vx} 
or the chiral quark-soliton model
($\chi$QSM)~\cite{Blotz:1992br,Blotz:1992pw}. 
Thus, some of model-dependent uncertainties are inexorable in
previous analyses of the SU(3) baryon masses. 

While the formalism of the baryon mass splittings was well established
within chiral soliton models, the numerical analyses are still
incomplete, because not all parameters can be fixed unequivocally, as
mentioned already. In order to avoid this ambiguity, it is essential
to consider the breakdown of isospin symmetry, without which it is
simply not possible to take as input the experimental data of the
octet baryon masses. The effects of isospin symmetry breaking consist
of two different contributions: The electromagnetic (EM) 
and hadronic ones. Those from the EM corrections were already
investigated in Ref.~\cite{Yang:2010id} within the same framework of
the $\chi$SM. Together with these EM corrections, the present authors
carried out a new analysis of the mass splittings of the SU(3)
baryons~\cite{Yang:2010fm}. Distinguished from the previous
works~\cite{Diakonov:1997mm,Ellis:2004uz}, all the dynamical
parameters were determined unambiguously, based on the experimental
baryon octet, $\Omega^-$ and $\Theta^+$ masses. We also showed that
the width of $\Theta^+$ and the transition magnetic moments for
$N^*(1685)\to N\gamma$ turned out to be consistent with the analysis
of the baryon mass splittings~\cite{Yang:2012cz}. Moreover, the
$\pi N$ sigma term, $\Sigma_{\pi N}$, was predicted to be
$\Sigma_{\pi N} = (36.4\pm 3.9)$ MeV. In the present work, we want to 
extend the investigation, taking into account the second-order
corrections of isospin and SU(3) symmetry breakings. It is important
to examine how stable the results with the first-order effects of
SU(3) and isospin symmetry breaking and how much we can improve the
numerical results in comparison with the existing experimental
data. For example, it will be shown that including the second-order
contributions leads to more consistent mass relations such as the
Morpugo mass formula~\cite{Morpurgo:1991if}.  

The present work is sketched as follows: In Section II, we recapitulate
all the formulae relevant to the mass splittings of the SU(3)
baryons. In Section III, We also discuss the results with 
the second-order corrections. In the last Section,
we summarize the present work and draw conclusions.

\section{Mass splittings of the SU(3) baryons from 
the chiral soliton model} 
The formalism for the mass splittings of the SU(3) baryons is
well known within chiral soliton models. In particular, the collective
Hamiltonian of chiral solitons have been investigated within various
versions of the $\chi$SM such as the Skyrme
model~\cite{Weigel:2008zz}, chiral quark-soliton
model~\cite{Blotz:1992pw,Blotz:1992br}, 
and chiral hyperbag model\cite{Park:1988vy}. We will recapitulate the
relevant formulae necessary for discussion of the baryon mass
splittings with the second-order SU(3) and isospin symmetry breakings.  
The collective Hamiltonian in the SU(3) $\chi\mathrm{SM}$ is expressed 
as 
\begin{eqnarray}
H &=& \frac{1}{2I_{1}}\sum_{i=1}^{3}\hat{J}_{i}^{2}
\;+\;\frac{1}{2I_{2}}\sum_{p=4}^{7}\hat{J}_{p}^{2} +
m_{\mathrm{ibr}}\left(\frac{\sqrt{3}}{2}\,\alpha\,
  D_{38}^{(8)}(A) 
\;+\;\beta\,\hat{T_{3}}
\;+\;\frac{1}{2}\,\gamma\sum_{i=1}^{3}D_{3i}^{(8)}(A)\,\hat{J}_{i}\right)\cr
&&
\;+\;m_{\mathrm{sbr}} \left(\alpha\,
   D_{88}^{(8)}(A)
\;+\;\beta\,\hat{Y}
\;+\;\frac{1}{\sqrt{3}}\,\gamma\sum_{i=1}^{3}D_{8i}^{(8)}(A)\,\hat{J}_{i}\right) 
\;+\;\left(m_{u}+m_{d}+m_{s}\right)\sigma,
\label{eq:sbH}
\end{eqnarray}
where $I_{1(2)}$ are the soliton moments of inertia that are dependent 
on dynamics of specific formulations of the $\chi$SM. The $\hat{J}_{i(p)}$
stand for the generators of the SU(3) group. The $\hat{Y}$ and
$\hat{T}_3$ denote the operators of the hypercharge and isospin third
component, respectively.  The $m_{\mathrm{u}}$,
$m_{\mathrm{d}}$, and $m_{\mathrm{s}}$ represent the up, down, and
strange current quark masses, respectively. $\bar{m}$,
$m_{\mathrm{ibr}}$, and $m_{\mathrm{sbr}}$ are defined respectively as 
\begin{equation}
\bar{m}\;=\; \frac{m_{\mathrm{u}}+m_{\mathrm{d}}}{2},\;\;\;\;
m_{\mathrm{ibr}} \;=\;  m_{\mathrm{d}} - m_{\mathrm{u}},\;\;\;\;
m_{\mathrm{sbr}} \;=\;  m_{\mathrm{s}} - \bar{m}.
\end{equation}
The $D_{ab}^{(\mathcal{R})}(A)$ are the SU(3) Wigner $D$
functions. The $\alpha$, $\beta$, and $\gamma$ encode dynamics of 
specific chiral soliton models. For example, in the chiral
quark-soliton model, they are written as 
\begin{equation}
\alpha\;=\;-\sigma+\frac{K_{2}}{I_{2}},
\;\;\;\;\beta\;=\;-\frac{K_{2}}{I_{2}},
\;\;\;\;\gamma\;=\;2\left(\frac{K_{1}}{I_{1}}-\frac{K_{2}}{I_{2}}\right),
\;\;\;\;\sigma\;\;=\;\;-(\alpha+\beta)\;=\;\frac{1}{3}\frac{\Sigma_{\pi 
    N}}{\bar{m}}.
\label{eq:abg}
\end{equation}

The eigenstates of the rotational part of the collective Hamiltonian,
i.e., of that without the symmetry breaking parts, 
are expressed in terms of the SU(3) Wigner $D$ functions in
representation $\mathcal{R}$:  
\begin{equation}
\Psi_{(\mathcal{R}^{*}\,;\, Y^{\prime}\, J\, J_{3})}^{(\mathcal{R\,};\, Y\, T\, T_{3})}(A)
\;=\;\sqrt{\textrm{dim}(\mathcal{R})}
\,(-)^{J_{3}+Y^{\prime}/2}\, D_{(Y,\, T,\, T_{3})(-Y^{\prime},\,
  J,\,-J_{3})}^{(\mathcal{R})*}(A),
\label{eq:Wigner}
\end{equation}
where $\mathcal{R}$ denotes the corresponding representation
of the $\mathrm{SU(3)}$ group, i.e., one of
$\mathcal{R}\,=\,8,\,10,\,\overline{10},\cdot\cdot\cdot$.  
$Y,\, T,\, T_{3}$ are the corresponding hypercharge, isospin,
and its third component, respectively. The constraint of the right
hypercharge $Y^{\prime}\;=\;1$ determines a tower of allowed
$\mathrm{SU(3)}$ representations: The lowest ones, that is, the baryon
octet and decuplet, coincide with those of the quark model. It
provides a certain duality between rigidly rotating heavy soliton and 
constituent quark model. The third lowest representation is the
antidecuplet~\cite{Diakonov:1997mm}. 

If one turns on SU(3) symmetry breaking, the collective wave functions
starts to get mixed with other representations\cite{Kim:1998gt}
\begin{eqnarray}
\left|B_{8}\right\rangle  & = & \left|8_{1/2},B\right\rangle 
\;+\; c_{\overline{10}}^{B}\left|\overline{10}_{1/2},B\right\rangle
\;+\; c_{27}^{B}\left|27_{1/2},B\right\rangle ,\cr
\left|B_{10}\right\rangle  & = & \left|10_{3/2},B\right\rangle 
\;+\; a_{27}^{B}\left|27_{3/2},B\right\rangle \;+\;
a_{35}^{B}\left|35_{3/2},B\right\rangle ,\cr
\left|B_{\overline{10}}\right\rangle  & = &
\left|\overline{10}_{1/2},B\right\rangle \;+\;
d_{8}^{B}\left|8_{1/2},B\right\rangle \;+\;
d_{27}^{B}\left|27_{1/2},B\right\rangle \;+\;
d_{\overline{35}}^{B}\left|\overline{35}_{1/2},B\right\rangle.
\label{eq:admix}
\end{eqnarray}
Here, the spin indices $J_{3}$ have been suppressed. The mixing
coefficients in Eq.(\ref{eq:admix}) read as follows
\begin{eqnarray}
c_{\overline{10}}^{B} & = & c_{\overline{10}}\left[\kern-0.5em 
\begin{array}{c}
\sqrt{5}\\
0\\
\sqrt{5}\\
0
\end{array}\kern-0.2em \right]\kern-0.2em ,\;\;\;
c_{27}^{B}=c_{27}\left[\kern-0.5em 
\begin{array}{c}
\sqrt{6}\\
3\\
2\\
\sqrt{6}
\end{array}\kern-0.2em \right]\kern-0.2em ,\;\;\;
a_{27}^{B}=a_{27}\left[\kern-0.5em 
\begin{array}{c}
\sqrt{15/2}\\
2\\
\sqrt{3/2}\\
0
\end{array}\kern-0.2em \right]\kern-0.2em ,\;\;\;
a_{35}^{B}=a_{35}\left[\kern-0.5em 
\begin{array}{c}
5/\sqrt{14}\\
2\sqrt{5/7}\\
3\sqrt{5/14}\\
2\sqrt{5/7}
\end{array}\kern-0.2em \right],\cr
d_{8}^{B} & = & d_{8}\left[
\begin{array}{c}
0\\
\sqrt{5}\\
\sqrt{5}\\
0
\end{array}\right],\;\;\; d_{27}^{B}=d_{27}\left[
\begin{array}{c}
0\\
\sqrt{3/10}\\
2/\sqrt{5}\\
\sqrt{3/2}
\end{array}\right],\;\;\;
d_{\overline{35}}^{B}=d_{\overline{35}}\left[
\begin{array}{c}
1/\sqrt{7}\\
3/(2\sqrt{14})\\
1/\sqrt{7}\\
\sqrt{5/56}
\end{array}\right],\label{eq:mix0}
\end{eqnarray}
respectively in the bases $[N,\;\Lambda,\;\Sigma,\;\Xi]$,
$[\Delta,\;\Sigma^{\ast},\;\Xi^{\ast},\;\Omega]$, $\left[\Theta^{+},\;
  N_{\overline{10}},\;\Sigma_{\overline{10}},\;\Xi_{\overline{10}}\right]$ 
and states in $\mathcal{R}=27,\;35,\;\overline{35}$. 
The coefficients in the mixing parameters are written in terms of
$\alpha$ and $\gamma$
\begin{eqnarray}
c_{\overline{10}} 
& = & -\frac{I_{2}}{15}m_{\mathrm{sbr}}
\left(\alpha+\frac{1}{2}\gamma\right),\;\;\;\;   
c_{27}=-\frac{I_{2}}{25}m_{\mathrm{sbr}}
\left(\alpha-\frac{1}{6}\gamma\right),\;\;\;\;
a_{27} \;=\; -\frac{I_{2}}{8}m_{\mathrm{sbr}}
\left(\alpha+\frac{5}{6}\gamma\right),\cr
a_{35} &=& -\frac{I_{2}}{24}m_{\mathrm{sbr}}
\left(\alpha-\frac{1}{2}\gamma\right),\;\;\;\;
d_{8} \;=\; \frac{I_{2}}{15}m_{\mathrm{sbr}}
\left(\alpha+\frac{1}{2}\gamma\right),\;\;\;\;
d_{27}\;=\;-\frac{I_{2}}{8}m_{\mathrm{sbr}}
\left(\alpha-\frac{7}{6}\gamma\right),\cr
d_{\overline{35}} &=& -\frac{I_{2}}{4}m_{\mathrm{sbr}}
\left(\alpha+\frac{1}{6}\gamma\right). 
\label{eq:mix1}
\end{eqnarray}

Here, we collect the expressions for the masses of the baryon octet,
decuplet, and antidecuplet. The first-order corrections of both isospin
and SU(3) symmetry breakings can be obtained as 
\begin{eqnarray}
M_{N}^{(1)} & = & 
m_{\mathrm{ibr}}\left(\frac{1}{10}\alpha+\beta-\frac{7}{20}\gamma\right)T_{3}
+m_{\mathrm{sbr}}\left(\frac{3}{10}\alpha+\beta-\frac{1}{20}\gamma\right),
\cr
M_{\Lambda}^{(1)} & = & 
m_{\mathrm{sbr}}\left(\frac{1}{10}\alpha+\frac{3}{20}\gamma\right),
\cr
M_{\Sigma}^{(1)} & = & 
m_{\mathrm{ibr}}\left(\frac{1}{4}\alpha+\beta-\frac{1}{8}\gamma\right)T_{3}
-m_{\mathrm{sbr}}\left(\frac{1}{10}\alpha+\frac{3}{20}\gamma\right),
\cr
M_{\Xi}^{(1)} & = & 
m_{\mathrm{ibr}}\left(\frac{2}{5}\alpha+\beta+\frac{1}{10}\gamma\right)T_{3}
-m_{\mathrm{sbr}}\left(\frac{1}{5}\alpha+\beta-\frac{1}{5}\gamma\right),
\cr
M_{\Delta}^{(1)} & = & 
m_{\mathrm{ibr}}\left(\frac{1}{8}\alpha+\beta-\frac{5}{16}\gamma\right)T_{3}
+m_{\mathrm{sbr}}\left(\frac{1}{8}\alpha+\beta-\frac{5}{16}\gamma\right),
\cr
M_{\Sigma^{\ast}}^{(1)} & = & 
m_{\mathrm{ibr}}\left(\frac{1}{8}\alpha+\beta-\frac{5}{16}\gamma\right)T_{3},
\cr
M_{\Xi^{\ast}}^{(1)} & = & 
m_{\mathrm{ibr}}\left(\frac{1}{8}\alpha+\beta-\frac{5}{16}\gamma\right)T_{3}
-m_{\mathrm{sbr}}\left(\frac{1}{8}\alpha+\beta-\frac{5}{16}\gamma\right),
\cr
M_{\Omega^{-}}^{(1)} & = & 
-2m_{\mathrm{sbr}}\left(\frac{1}{8}\alpha+\beta-\frac{5}{16}\gamma\right),
\cr
M_{\Theta^{+}}^{(1)} & = & 
2m_{\mathrm{sbr}}\left(\frac{1}{8}\alpha+\beta-\frac{1}{16}\gamma\right),
\cr
M_{N^{\ast}}^{(1)} & = & 
m_{\mathrm{ibr}}\left(\frac{1}{8}\alpha+\beta-\frac{1}{16}\gamma\right)T_{3}
+m_{\mathrm{sbr}}\left(\frac{1}{8}\alpha+\beta-\frac{1}{16}\gamma\right),
\cr
M_{\Sigma_{\overline{10}}}^{(1)} & = & 
m_{\mathrm{ibr}}\left(\frac{1}{8}\alpha+\beta-\frac{1}{16}\gamma\right)T_{3},
\cr
M_{\Xi_{3/2}}^{(1)} & = & 
m_{\mathrm{ibr}}\left(\frac{1}{8}\alpha+\beta-\frac{1}{16}\gamma\right)T_{3}
-m_{\mathrm{sbr}}\left(\frac{1}{8}\alpha+\beta-\frac{1}{16}\gamma\right).
\label{eq:firstM}
\end{eqnarray}

The second-order corrections to the masses of the SU(3) baryons can
be derived perturbatively~\cite{Ellis:2004uz}: 
\begin{equation}
M_{B_{\nu}}^{(2)}
\;=\;-\sum_{n\neq\nu}\frac{|\langle
  B_{n}|H_{\mathrm{sb}}|B_{\nu}\rangle|^{2}}
{\Delta M_{n-\nu}^{(0)}},
\end{equation}
where $\Delta M_{n\neq\nu}^{(0)}$ denotes the differences between
the centers of the multiplets~\cite{Diakonov:1997mm}.
The second-order corrections are given as 
\begin{eqnarray}
M_{N}^{(2)} & = & -I_{2}\,
 T_{3}^{2}\left[\frac{1}{750}m_{\mathrm{ibr}}^{2}
   \left(25\left(\alpha-\frac{1}{6}\gamma\right)^{2}  
+2\left(\alpha+\frac{1}{18}\gamma\right)^{2}\right) 
+\;\frac{2}{375}m_{\mathrm{sbr}}^{2}
\left(18\left(\alpha-\frac{1}{6}\gamma\right)^{2}
+25\left(\alpha+\frac{1}{2}\gamma\right)^{2}\right)\right],\cr
M_{\Lambda}^{(2)} & = &
 -\frac{9}{250}I_{2}m_{\mathrm{sbr}}^{2}
\left(\alpha-\frac{1}{6}\gamma\right)^{2},\cr
M_{\Sigma}^{(2)} & = & 
-I_{2}\left[\frac{1}{120}m_{\mathrm{ibr}}^{2}
\left(\alpha-\frac{1}{6}\gamma\right)^{2}T_{3}^{2}
\;+\;\frac{1}{750}m_{\mathrm{sbr}}^{2}
\left(12\left(\alpha-\frac{1}{6}\gamma\right)^{2}
+25\left(\alpha+\frac{1}{2}\gamma\right)^{2}\right)\right],\cr
M_{\Xi}^{(2)} & = & 
-I_{2}\, 
 T_{3}^{2}\left[ m_{\mathrm{ibr}}^2
\frac{1}{375}\left(\alpha+\frac{1}{18}\gamma\right)^{2}
\;+\;\frac{12}{125}m_{\mathrm{sbr}}^{2}
\left(\alpha-\frac{1}{6}\gamma\right)^{2}\right],\cr
M_{\Delta}^{(2)} & = & 
-I_{2}\left[m_{\mathrm{ibr}}^{2}\left(\frac{1}{2688}
\left(\alpha-\frac{1}{2}\gamma\right)^{2}+\frac{5}{384}
\left(\alpha+\frac{5}{6}\gamma\right)^{2}\right)T_{3}^{2}
+m_{\mathrm{sbr}}^{2}\left(\frac{25}{2688}
\left(\alpha-\frac{1}{2}\gamma\right)^{2}
+\frac{15}{128}\left(\alpha+\frac{5}{6}\gamma\right)^{2}\right)\right],\cr
M_{\Sigma^{\ast}}^{(2)} & = & 
-I_{2}\left[m_{\mathrm{ibr}}^{2}\left(\frac{5}{5376}
\left(\alpha-\frac{1}{2}\gamma\right)^{2}+\frac{9}{256}
\left(\alpha+\frac{5}{6}\gamma\right)^{2}\right)T_{3}^{2}
+m_{\mathrm{sbr}}^{2}\left(\frac{5}{336}
\left(\alpha-\frac{1}{2}\gamma\right)^{2}+\frac{1}{16}
\left(\alpha+\frac{5}{6}\gamma\right)^{2}\right)\right],\cr
M_{\Xi^{\ast}}^{(2)} & = & 
-I_{2}\left[m_{\mathrm{ibr}}^{2}\left(\frac{5}{21504}
\left(\alpha-\frac{1}{2}\gamma\right)^{2}+\frac{49}{3072}
\left(\alpha+\frac{5}{6}\gamma\right)^{2}\right)
+ m_{\mathrm{ibr}}^{2}\left(\frac{5}{5376}
\left(\alpha-\frac{1}{2}\gamma\right)^{2}+\frac{49}{768}
\left(\alpha+\frac{5}{6}\gamma\right)^{2}\right)T_{3}^{2}\right. \cr
 &  & \hspace{0.5cm}\left.+\;m_{\mathrm{sbr}}^{2}\left(\frac{15}{896}
\left(\alpha-\frac{1}{2}\gamma\right)^{2}+\frac{3}{128}
\left(\alpha+\frac{5}{6}\gamma\right)^{2}\right)\right],\cr
M_{\Omega^{-}}^{(2)} & = &
 -I_{2}\frac{5}{336}(m_{\mathrm{s}}-\bar{m})^{2}
\left(\alpha-\frac{1}{2}\gamma\right)^{2},\cr
M_{\Theta^{+}}^{(2)} & = & -\frac{3}{112}I_{2}
m_{\mathrm{sbr}}^{2}\left(\alpha+\frac{1}{6}\gamma\right)^{2},\cr
M_{N^{\ast}}^{(2)} & = &
 -I_{2}\left[m_{\mathrm{ibr}}^{2}\left(\frac{3}{896}
\left(\alpha-\frac{1}{18}\gamma\right)^{2}\;-\;\frac{1}{30}
\left(\alpha-\frac{1}{6}\gamma\right)^{2}+\;\frac{49}{1920}
\left(\alpha+\frac{7}{18}\gamma\right)^{2}\right)T_{3}^{2}\right.\cr
 & & \hspace{0.5cm} \left.+\;m_{\mathrm{sbr}}^{2}\left(\frac{3}{640}
\left(\alpha-\frac{7}{6}\gamma\right)^{2}+\frac{27}{896}
\left(\alpha+\frac{1}{6}\gamma\right)^{2}-\frac{1}{30}
\left(\alpha+\frac{1}{2}\gamma\right)^{2}\right)\right],\cr
M_{\Sigma_{\overline{10}}}^{(2)} & = &
 -I_{2}\left[m_{\mathrm{ibr}}^{2}\left(\frac{3}{1792}
\left(\alpha-\frac{1}{18}\gamma\right)^{2}\;-\;\frac{1}{120}
\left(\alpha-\frac{1}{6}\gamma\right)^{2}-\frac{9}{1280}
\left(\alpha+\frac{7}{18}\gamma\right)^{2}\right)T_{3}^{2}\right.\cr
 &  & \hspace{0.5cm}\left. +\; m_{\mathrm{sbr}}^{2}\left(\frac{1}{80}
\left(\alpha-\frac{7}{6}\gamma\right)^{2}+\frac{3}{112}
\left(\alpha+\frac{1}{6}\gamma\right)^{2}-\frac{1}{30}
\left(\alpha+\frac{1}{2}\gamma\right)^{2}\right)\right],\cr
M_{\Xi_{\overline{10}}}^{(2)} & = &
 -I_{2}\left[m_{\mathrm{ibr}}^{2}
\left(\frac{3}{4480}\left(\alpha-\frac{1}{18}\gamma\right)^{2}
+\frac{1}{384}\left(\alpha+\frac{7}{18}\gamma\right)^{2}\right)T_{3}^{2}\right. \cr 
&& \hspace{0.5cm}\left. +\;m_{\mathrm{sbr}}^{2}\left(\frac{3}{128}
\left(\alpha-\frac{7}{6}\gamma\right)^{2}+\frac{15}{896}
\left(\alpha+\frac{1}{6}\gamma\right)^{2}\right)\right].
\label{eq:2nd_M}
\end{eqnarray}
The masses of the SU(3) baryons can be expressed in terms of $M_{B}^{(1)}$
and $M_{B}^{(2)}$: 
\begin{equation}
M_{B}\;=\;\overline{M}_{B}+M_{B}^{(1)}+M_{B}^{(2)},
\end{equation}
where $\overline{M}_{B}$ stand for the center masses of the multiplets.

The EM mass corrections to SU(3) baryon masses were already discussed
in Ref.~\cite{Yang:2010id}. We compile the corresponding formulae
here for the baryon octet 
\begin{eqnarray}
M_{N}^{\mathrm{EM}} & = &
\frac{1}{5}\left(c^{(8)}+\frac{4}{9}c^{(27)}\right)T_{3}
+\frac{3}{5}\left(c^{(8)}+\frac{2}{27}c^{(27)}\right)\left(T_{3}^{2}
+\frac{1}{4}\right)+c^{(1)},\cr
M_{\Lambda}^{\mathrm{EM}} & = & \frac{1}{10}\left(c^{(8)}
-\frac{2}{3}c^{(27)}\right)+c^{(1)},\cr
M_{\Sigma}^{\mathrm{EM}} & = & \frac{1}{2}c^{(8)}\, T_{3}
+\frac{2}{9}c^{(27)}\, T_{3}^{2}-\frac{1}{10}\left(c^{(8)}
+\frac{14}{9}c^{(27)}\right)+c^{(1)},\cr
M_{\Xi}^{\mathrm{EM}} & = & \frac{4}{5}\left(c^{(8)}
-\frac{1}{9}c^{(27)}\right)T_{3}-\frac{2}{5}\left(c^{(8)}
-\frac{1}{9}c^{(27)}\right)\left(T_{3}^{2}+\frac{1}{4}\right)
+c^{(1)},
\label{eq:emforoc}
\end{eqnarray}
and for the baryon decuplet 
\begin{eqnarray}
M_{\Delta}^{\mathrm{EM}} 
& = & \frac{1}{4}\left(c^{(8)}+\frac{8}{63}c^{(27)}\right)T_{3}+
\frac{5}{63}c^{(27)}\, T_{3}^{2}
+\frac{1}{8}\left(c^{(8)}-\frac{2}{3}c^{(27)}\right)+c^{(1)},\cr
M_{\Sigma^{\ast}}^{\mathrm{EM}} 
& = & \frac{1}{4}\left(c^{(8)}-\frac{4}{21}c^{(27)}\right)T_{3}
+\frac{5}{63}c^{(27)}\,\left(T_{3}^{2}-1\right)+c^{(1)},\cr
M_{\Xi^{\ast}}^{\mathrm{EM}} 
& = & \frac{1}{4}\left(c^{(8)}-\frac{32}{63}c^{(27)}\right)T_{3}
-\frac{1}{4}\left(c^{(8)}
+\frac{8}{63}c^{(27)}\right)\left(T_{3}^{2}+\frac{1}{4}\right)+c^{(1)},\cr
M_{\Omega^{-}}^{\mathrm{EM}} 
& = & -\frac{1}{4}\left(c^{(8)}-\frac{4}{21}c^{(27)}\right)+c^{(1)},
\label{eq:emfordec}
\end{eqnarray}
and for the baryon antidecuplet
\begin{eqnarray}
M_{\Theta^{+}}^{\mathrm{EM}} & = & 
\frac{1}{4}\left(c^{(8)}-\frac{4}{21}c^{(27)}\right)+c^{(1)},\cr
M_{N^{\ast}}^{\mathrm{EM}} & = & 
\frac{1}{4}\left(c^{(8)}-\frac{32}{63}c^{(27)}\right)T_{3}
+\frac{1}{4}\left(c^{(8)}
+\frac{8}{63}c^{(27)}\right)\left(T_{3}^{2}+\frac{1}{4}\right)+c^{(1)},\cr
M_{\Sigma_{\overline{10}}}^{\mathrm{EM}} & = & 
\frac{1}{4}\left(c^{(8)}-\frac{4}{21}c^{(27)}\right)T_{3}
-\frac{5}{63}c^{(27)}\left(T_{3}^{2}-1\right)+c^{(1)},\cr
M_{\Xi_{3/2}^{+}}^{\mathrm{EM}} & = & 
\frac{1}{4}\left(c^{(8)}+\frac{8}{63}c^{(27)}\right)T_{3}
-\frac{5}{63}c^{(27)}T_{3}^{2}-\frac{1}{8}\left(c^{(8)}
-\frac{2}{3}c^{(27)}\right)+c^{(1)},
\label{eq:antidecM_EM}
\end{eqnarray}
respectively.

Since the center of baryon masses absorb the singlet contributions
to the EM masses with $c^{(1)}$, we safely neglect them for EM
mass differences. At any rate, they are not pertinent to the EM mass
differences in which they are canceled out. Therefore, the expressions
of EM mass differences of SU(3) baryons have only two unknown parameters,
i.e., $c^{(8)}$ and $c^{(27)}$, which were found to be
\begin{equation}
c^{(8)}\;=\;-0.15\pm0.23,\;\;\;\;\;\; c^{(27)}\;=\;8.62\pm2.39,
\label{eq:c8c27}
\end{equation}
in units of MeV~\cite{Yang:2010id}.

With SU(3) symmetry and isospin symmetry breakings considered, the
masses of the SU(3) baryons can be expressed in terms of
all the contributions discussed above
\begin{equation}
M_{B}\;=\;\overline{M}_{B}+M_{B}^{(1)}+M_{B}^{(2)}+M_{B}^{\mathrm{EM}}.
\end{equation}

\section{Results and discussion}
The mass splittings of the SU(3) baryons with SU(3) and isospin
symmetry breakings to the first order were already investigated in
Ref.~\cite{Yang:2010fm}.  Thus, we will show in this Section how to
analyze the mass splittings of SU(3) baryons, considering the
second-order corrections of isospin and SU(3) 
flavor symmetry breakings. Though the second-order effects of isospin
symmetry breaking are rather small, we will take into account those
effects, first for a consistency reason, and second for their practical
importance. We will soon see that the results are more 
consistent with the existing experimental data. Though the 
second-order corrections of SU(3) flavor symmetry breaking to the
mass splittings of the SU(3) baryon were already studied in
Ref.~\cite{Ellis:2004uz}, it is still incomplete, because it is not
possible to use the baryon octet masses as input without isospin
symmetry breaking. Furthermore, those effects of isospin symmetry
breaking come into play in improving the mass splittings within the
isospin multiplets. 

The general method for determining all model 
parameters is very similar to the case of the first-order
analysis~\cite{Yang:2010fm}. We employ the least-square fit to adjust
the model parameters, using as input the masses of the whole baryon
octet, $\Omega^-$, and $\Theta^+$ from the LEPS experiment.   
In order to determine the masses of the decuplet and antidecuplet, we
have to know at least the mass of one member in each representation. 
We have selected $\Omega^-$ and $\Theta^+$, because they are the
isospin singlet in the decuplet and antidecuplet representations,
respectively such that we can fix the center masses uniquely. 
One could choose other sets of baryons but they would yield
the results that are phenomenologically worse.

\begin{table}[htp]
\caption{The comparison of important parameters from the 1st order
  mass corrections with those from the full (to 2nd-order) ones. Input
  values of octet and decuplet baryon masses are taken from
  experimental data \cite{PDG}.} 

\begin{tabular}{c|c|c}
\hline 
 & First-order results  & Full results \tabularnewline
\hline 
$I_{2}$  & $0.420\pm0.006\;\mathrm{fm}$  
& $0.431\pm0.001\;\mathrm{fm}$ \tabularnewline
$m_{\mathrm{ibr}}\alpha$  
& $-4.390\pm0.004\;\mathrm{MeV}$ 
& $-6.458\pm0.004\;\mathrm{MeV}$ \tabularnewline
$m_{\mathrm{ibr}}\beta$  
& $-2.411\pm0.001\;\mathrm{MeV}$ 
& $-2.972\pm0.001\;\mathrm{MeV}$ \tabularnewline
$m_{\mathrm{ibr}}\gamma$  
& $-1.740\pm0.006\;\mathrm{MeV}$ 
& $-2.288\pm0.008\;\mathrm{MeV}$ \tabularnewline
$m_{\mathrm{sbr}}\alpha$  
& $-255.029\pm5.821\;\mathrm{MeV}$ 
& $-280.8\pm14.2\;\mathrm{MeV}$ \tabularnewline
$m_{\mathrm{sbr}}\beta$  
& $-140.040\pm3.195\;\mathrm{MeV}$ 
& $-129.3\pm6.5\;\mathrm{MeV}$ \tabularnewline
$m_{\mathrm{sbr}}\gamma$  
& $-101.081\pm2.332\;\mathrm{MeV}$ 
& $-99.5\pm5.0\;\mathrm{MeV}$ \tabularnewline
$\Sigma_{\pi N}$  
& $36.4\pm3.9\;\mathrm{MeV}$ 
& $50.5\pm5.4\;\mathrm{MeV}$ \tabularnewline
$c_{\overline{10}}$  
& $0.0434\pm0.0006$ 
& $0.0482\pm0.0021$ \tabularnewline
$c_{27}$  & $0.0203\pm0.0003$ 
& $0.0231\pm0.0012$ \tabularnewline
$a_{27}$  & $0.0903\pm0.0013$ 
& $0.0994\pm0.0004$ \tabularnewline
$a_{35}$  & $0.0181\pm0.0003$ 
& $0.0210\pm0.0013$ \tabularnewline
$d_{8}$  & $-0.0434\pm0.0006$ 
& $-0.0482\pm0.0021$ \tabularnewline
$d_{27}$  & $0.0365\pm0.0005$ 
& $0.0450\pm0.0042$ \tabularnewline
$d_{\overline{35}}$  
& $0.1447\pm0.0021$ 
& $0.1625\pm0.0078$ \tabularnewline
\hline 
\end{tabular}\label{tab:comp-2} 
\end{table}
The results of the fixed parameters are listed in Table~\ref{tab:comp-2}
in comparison with those from the analysis with the first-order corrections
only. As shown in Table~\ref{tab:comp-2}, the parameters with the
second-order corrections are changed from those with the first-order
ones. Almost all parameters are altered by about $(20-30)\,\%$. 
In particular, the $\pi N$ sigma term turns out to be $(50.5\pm5.4)$
MeV, which is almost $30\,\%$ larger than that with the first-order
corrections ($(36.4\pm3.9)$ MeV). This can be easily understood from
Eq.~(\ref{eq:abg}) in which the $\pi N$ sigma term is expressed as
$\alpha+\beta$. In fact, this sum of $\alpha$ and $\gamma$ is enhanced
in magnitude with the second-order corrections.
Consequently, the $\pi N$ sigma term is increased by about $30\,\%$
with the second-order corrections taken into account. In
Ref.~\cite{Diakonov:1997mm}, $\Sigma_{\pi N}=45\,\mathrm{MeV}$ was
used~\cite{Gasser:1990ce}, while Ref.~\cite{Ellis:2004uz} obtained 
$\Sigma_{\pi N}=73\,\mathrm{MeV}$ in studying the baryon
antidecuplet~\cite{Pavan:2001wz,Schweitzer:2003fg}. 
It was discussed in Ref.~\cite{Schweitzer:2003fg} that larger values
of the $\Sigma_{\pi N}$ are preferable to describe the then mass
splitting in the baryon antidecuplet, because the debatable NA49 data
of the $\Xi_{3/2}$ mass~\cite{Alt:2003vb} was used for determining
$\Sigma_{\pi N}$. Indeed, the larger value of the $\Sigma_{\pi N}$
reduces the antidecuplet splitting noticeably~\cite{Diakonov:2003jj}.
In Ref.~\cite{Schweitzer:2003fg}, the $\Sigma_{\pi N}$ has been
extracted by using the $\Theta^{+}$ and $\Xi_{3/2}$ masses, based
on the $\chi$QSM: $\Sigma_{\pi N}=(74\pm12)\,\mathrm{MeV}$, which is
quite larger than the present value. As we will show later, the
predicted mass of $\Xi_{3/2}$ is larger than the NA49 data.  

We are now in a position to present the main results of the present
work, i.e., the masses of the SU(3) baryons. In Table~\ref{tab:Octm1}
we list the reproduced masses of the baryon octet. The results indicate
the stability of the numerical analysis. The sixth and seventh columns
in Table~\ref{tab:Octm1} represent the reproduced octet masses respectively
with the first-order and full contributions of isospin and SU(3) flavor
symmetry breakings taken into account. The results with the full contributions
are shown mostly to be closer to the input masses, as expected. 
\begin{table}[htp]
\centering \caption{Reproduced masses of the baryon octet. The
  experimental data of octet are taken from the Particle Data Group
  (PDG)~\cite{PDG}.  }

\begin{tabular}{ccccccc}
\hline 
\multicolumn{2}{c}{Mass {[}MeV{]}} 
& $T_{3}$  
& $Y$  & Exp. {[}input{]}  
& $M_{B_{8}}\,(\mbox{to 1st order})$  
& $M_{B_{8}}\,(\mbox{full})$\tabularnewline
\hline 
$M_{N}$  & $\begin{array}{c}
p\\
n
\end{array}$  & $\begin{array}{c}
\;\;1/2\\
-1/2
\end{array}$  & $\;\;1$  & $\begin{array}{c}
938.27203\pm0.00008\\
939.56536\pm0.00008
\end{array}$  & $\begin{array}{c}
939.8\pm3.7\\
940.3\pm3.6
\end{array}$  & $\begin{array}{c}
938.2\pm8.5\\
940.3\pm8.5
\end{array}$\tabularnewline
\hline 
$M_{\Lambda}$  
& $\Lambda$  
& $\;\;0$  
& $\;\;0$  
& $1115.683\pm0.006$  
& $1109.6\pm0.7$  
& $1110.2\pm2.1$ \tabularnewline
\hline 
$M_{\Sigma}$  & $\begin{array}{c}
\Sigma^{+}\\
\Sigma^{0}\\
\Sigma^{-}
\end{array}$  & $\begin{array}{c}
\;\;1\\
\;\;0\\
-1
\end{array}$  
& $\;\;0$  
& $\begin{array}{c}
1189.37\;\pm0.07\\
1192.642\pm0.024\\
1197.449\pm0.030
\end{array}$  & $\begin{array}{c}
1188.8\pm0.7\\
1190.2\pm0.8\\
1195.5\pm0.7
\end{array}$  & $\begin{array}{c}
1188.1\pm0.8\\
1190.6\pm0.9\\
1196.9\pm0.8
\end{array}$\tabularnewline
\hline 
$M_{\Xi}$  & $\begin{array}{c}
\Xi^{0}\\
\Xi^{-}
\end{array}$  & $\begin{array}{c}
\;\;1/2\\
-1/2
\end{array}$  & $-1$  & $\begin{array}{c}
1314.83\pm0.20\\
1321.31\pm0.13
\end{array}$  & $\begin{array}{c}
1319.3\pm3.4\\
1324.5\pm3.4
\end{array}$  & $\begin{array}{c}
1318.1\pm7.1\\
1324.8\pm7.1
\end{array}$\tabularnewline
\hline 
\end{tabular}\label{tab:Octm1} 
\end{table}

\begin{table}[htp]
\centering \caption{The results of the masses of the baryon decuplet. The experimental
data are taken from the Particle Data Group (PDG)~\cite{PDG}. }

\begin{tabular}{ccccccc}
\hline 
\multicolumn{2}{c}{Mass {[}MeV{]}} 
& $T_{3}$  
& $Y$  & Experiment \cite{PDG}  
& $M_{B_{10}}\,(\mbox{to 1st order})$  
& $M_{B_{10}}\,(\mbox{full})$ \tabularnewline
\hline 
$M_{\Delta}$  & $\begin{array}{c}
\Delta^{++}\\
\Delta^{+}\\
\Delta^{0}\\
\Delta^{-}
\end{array}$  & $\begin{array}{c}
\;\;3/2\\
\;\;1/2\\
-1/2\\
-3/2
\end{array}$  & $\;\;1$  
& $1231-1233$  & $\begin{array}{c}
1248.5\pm3.4\\
1249.4\pm3.4\\
1251.5\pm3.4\\
1255.1\pm3.4
\end{array}$  & $\begin{array}{c}
1235.4\pm8.0\\
1236.8\pm8.0\\
1239.7\pm8.0\\
1243.9\pm8.0
\end{array}$ \tabularnewline
\hline 
$M_{\Sigma^{\ast}}$  & $\begin{array}{c}
\Sigma^{\ast+}\\
\Sigma^{\ast0}\\
\Sigma^{\ast-}
\end{array}$  & $\begin{array}{c}
\;\;1\\
\;\;0\\
-1
\end{array}$  & $\;\;0$  & $\begin{array}{c}
1382.8\pm0.4\\
1383.7\pm1.0\\
1387.2\pm0.5
\end{array}$  & $\begin{array}{c}
1388.5\pm0.3\\
1390.7\pm0.4\\
1394.2\pm0.3
\end{array}$  & $\begin{array}{c}
1383.8\pm1.7\\
1386.6\pm1.7\\
1390.8\pm1.7
\end{array}$ \tabularnewline
\hline 
$M_{\Xi^{\ast}}$  & $\begin{array}{c}
\Xi^{\ast0}\\
\Xi^{\ast-}
\end{array}$  & $\begin{array}{c}
\;\;1/2\\
-1/2
\end{array}$  & $-1$  & $\begin{array}{c}
1531.80\pm0.32\\
1535.0\;\pm0.6\;\,
\end{array}$  & $\begin{array}{c}
1529.8\pm3.4\\
1533.3\pm3.4
\end{array}$  & $\begin{array}{c}
1529.5\pm6.8\\
1533.7\pm6.8
\end{array}$\tabularnewline
\hline 
$M_{\Omega^{-}}$  & $\Omega^{-}$  
& $0$  & $-2$  & $1672.45\pm0.29$  
& input  & input\tabularnewline
\hline 
\end{tabular}\label{tab:decm1} 
\end{table}

In Table~\ref{tab:decm1}, we list the results of the masses of the 
baryon decuplet. The last column represents the final results of the
present work with the full contributions considered. As shown in
Table~~\ref{tab:decm1}, they are in better agreement with the
experimental data in general, compared to those with the first-order
corrections only (in the fifth column). The masses of the baryon
decuplet are in general well reproduced. 

\begin{table}[htp]
\centering \caption{The results of the masses of the baryon antidecuplet. }

\begin{tabular}{ccccccc}
\hline 
\multicolumn{2}{c}{Mass } & $T_{3}$  
& $Y$  & Experiment  & $M_{B_{\overline{10}}}\,(\mbox{to 1st order})$  
& $M_{B_{\overline{10}}}\,(\mbox{full})$ \tabularnewline
\hline 
$M_{\Theta^{+}}$  & $\Theta^{+}$  
& $\;\;0$  & $\;\;2$  & $1524\pm0.005$\cite{Nakano:2008ee}  
& input  & input\tabularnewline
\hline 
$M_{N^{\ast}}$  & $\begin{array}{c}
p^{\ast}\\
n^{\ast}
\end{array}$  & $\begin{array}{c}
\;\;1/2\\
-1/2
\end{array}$  & $\;\;1$  
& ${\displaystyle 1685\pm0.012}$\cite{Kuznetsov:2007dy}  
& $\begin{array}{c}
1688.2\pm10.5\\
1692.2\pm10.5
\end{array}$  & $\begin{array}{c}
1687.4\pm6.8\\
1692.2\pm6.8
\end{array}$\tabularnewline
\hline 
$M_{\Sigma_{\overline{10}}}$  & $\begin{array}{c}
\Sigma_{\overline{10}}^{+}\\
\Sigma_{\overline{10}}^{0}\\
\Sigma_{\overline{10}}^{-}
\end{array}$  & $\begin{array}{c}
\;\;1\\
\;\;0\\
-1
\end{array}$  & $\;\;0$  &  & $\begin{array}{c}
1852.4\pm10.0\\
1856.3\pm10.0\\
1859.0\pm10.0
\end{array}$  & $\begin{array}{c}
1844.0\pm0.2\\
1848.7\pm0.3\\
1852.1\pm0.2
\end{array}$\tabularnewline
\hline 
$M_{\Xi_{3/2}}$  & $\begin{array}{c}
\Xi_{3/2}^{+}\\
\Xi_{3/2}^{0}\\
\Xi_{3/2}^{-}\\
\Xi_{3/2}^{--}
\end{array}$  & $\begin{array}{c}
\;\;3/2\\
\;\;1/2\\
-1/2\\
-3/2
\end{array}$  & $\;\;-1$  
&  & $\begin{array}{c}
2016.5\pm10.5\\
2020.5\pm10.5\\
2023.1\pm10.5\\
2024.4\pm10.5
\end{array}$  & $\begin{array}{c}
1993.7\pm6.7\\
1998.5\pm6.7\\
2001.9\pm6.7\\
2003.9\pm6.7
\end{array}$\tabularnewline
\hline 
\end{tabular}\label{tab:antim1} 
\end{table}
Table~\ref{tab:antim1} presents the results of the masses of the baryon
antidecuplet. Note that the mass of the $\Theta^{+}$ is used as input.
Since there are not enough experimental data, we can only compare
the results of $N^{*}(1685)$ with the experimental data and find
that they are in agreement with the data. The nature of this $N^{*}(1685)$
resonance is not reached yet in consensus. For example,
Ref.~\cite{Shklyar:2006xw} suggests that this resonance arises from
coupled channel effects of the $\mathrm{S}_{11}(1535)$,
$\mathrm{S}_{11}(1650)$, and $P_{11}(1710)$ resonances. However, the
present analysis identifies it preferably as a member of the baryon
antidecuplet. 

As mentioned previously, it is known that the larger value of the
$\Sigma_{\pi N}$ reduces the antidecuplet splitting noticeably
~\cite{Diakonov:2003jj,Schweitzer:2003fg,Ellis:2004uz}.
However, the present result of $\Sigma_{\pi N}$ turns out
to be smaller than that found in Ref.~\cite{Ellis:2004uz}, 
even though we have considered the second-order
contributions of isospin and SU(3) flavor symmetry breakings. 
Moreover, the mass splittings of the baryon antidecuplet remain rather
stable with the second-order corrections. As discussed before, the
sigma $\pi N$ term is proportional 
to $m_{\mathrm{ibr}}\sigma$ in Eq.~(\ref{eq:abg}). While the change
of $m_{\mathrm{ibr}}\sigma$ affects $\Sigma_{\pi N}$, the isospin mass
splittings are rather insensitive to the second-order corrections
of isospin symmetry breaking. As a result, the mass splittings of
the baryon antidecuplet remain stable. Note that the masses of $\Xi_{3/2}$
are found to be larger than the NA49 data~\cite{Alt:2003vb},
though its existence is under debate. In Ref.~\cite{Goeke:2009ae} 
the mass ranges of the $\Sigma_{\overline{10}}$ and $\Xi_{3/2}$
were derived as $1795<\Sigma_{\overline{10}}<1830$ MeV and
$1900<\Xi_{3/2}<1970$ MeV. The present results turn out to be slightly
larger than those of Ref.~\cite{Goeke:2009ae}.

Finally, it is interesting to mention that Ref.~\cite{Morpurgo:1991if}
defined a quantity measuring the strength of the second-order
corrections of SU(3) symmetry breaking as 
\begin{equation}
T\;=\;
M_{\Xi^{\ast-}}-\frac{1}{2}\left(M_{\Sigma^{\ast-}}+M_{\Omega^{-}}\right),
\label{eq:morpugo}
\end{equation}
which vanishes at the level of the first-order corrections, as pointed
out in Ref.~\cite{Morpurgo:1991if}. It is indeed so, since it becomes
nonzero only when the second-order corrections are considered as
follows: 
\begin{equation}
T\;=\; m_{\mathrm{sbr}}^2\,
I_{2}\left[\frac{1}{128}\left(\alpha+\frac{5\gamma}{6}\right)^{2}
-\frac{5}{2688}\left(\alpha-\frac{\gamma}{2}\right)^{2}\right]
\;=\;(2.04\pm0.16)\,\mathrm{MeV}.
\label{eq:nonzero}
\end{equation}
The present value of $T$ seems smaller than the experimental
one $(5.2\pm1.3)\,\mathrm{MeV}$.

\section{Summary and conclusion}
In the present work, we have investigated the masses of the SU(3)
baryons within the framework of a chiral soliton model, taking into
account SU(3) and isospin symmetry breakings to the second order in
the perturbative expansion of the current quark masses. We also have
considered the electromagnetic self-interactions that contribute to
the isospin mass splittings. In order to determine the unknown model
parameters $\alpha$, $\beta$, and $\gamma$, we employed the
experimental data of the baryon octet, the $\Omega^{-}$, and the
$\Theta^{+}$. We then performed the minimization of the
$\chi^{2}$. The second moment of inertia $I_{2}$ was also found, which
is a key parameter to explain the mass splittings within the baryon
antidecuplet. Moreover, the pion-nucleon sigma term was determined to
be $\Sigma_{\pi N}=(50.5\pm5.4)$ MeV. The present results of the
baryon decuplet masses are in remarkable agreement with the
experimental data. The masses of $N^{*}(1685)$ in the antidecuplet
turned out to be very close to the recent experimental 
data.

The present work is distinguished from the previous
studies~\cite{Diakonov:1997mm,Ellis:2004uz} 
based on the chiral soliton model, which also deal with the mass splittings
of the SU(3) baryons. The second moment of inertia $I_{2}$ plays
a crucial role in explaining the heavier masses of the baryon antidecuplet,
compared to those of the octet and decuplet. However, it was not possible
to fix it unambiguously in previous works. In particular, since the
$\Sigma_{\pi N}$ was not uniquely known empirically, some ambiguities
were inevitable in previous analyese. While
Refs.~\cite{Diakonov:1997mm,Ellis:2004uz} 
used the experimental data for the baryon octet, they did not consider
isospin symmetry breaking, so that they were unable to incorporate
whole experimental information. On the other hand, we were able to
fix all model parameters by using the experimental data for the masses
of the baryon octet, $\Omega^{-}$, and $\Theta^{+}$, because effects
of isospin symmetry breaking (both hadronic and electromagnetic parts)
have been fully taken into account. Thus, we have produced the masses
of the baryon antidecuplet as well as of the decuplet without any
further adjustable parameter.

The vector and axial-vector properties of the SU(3) baryons can be
investigated in a similar ``model-independent'' analysis. However,
the previous analyses also suffer from ambiguities in determining
parameters~\cite{Diakonov:2003jj,Ellis:2004uz,Kim:1997ip,Kim:1999uf,
Yang:2007yj}. 
The parameters fixed within this work can be used in determining the
magnetic moments and axial-vector constants of the SU(3) baryons.
The related works are in progress~\cite{YangKim}. 
\begin{acknowledgments}
The present work was supported by Basic Science Research Program
through the National Research Foundation of Korea funded by the
Ministry of Education, Science and Technology (Grant Number:
2009-0089525). 
\end{acknowledgments}
%\cite{Jezabek:1987ns}

\end{document}